# Discrete solitons and soliton-induced dislocations in partially-coherent photonic lattices

**Hector Martin[1], Eugenia D. Eugenieva[2] and Zhigang Chen[1,3]**

*[1] Department of Physics and Astronomy, San Francisco State University, CA 94132*
*[2] Intel Corporation, USA, [3] TEDA College, Nankai University, China*
*zchen@stars.sfsu.edu*

**Demetrios N. Christodoulides**

*School of Optics/CREOL, University of Central Florida, Orlando, FL 32816*

**Abstract:** We investigate both experimentally and theoretically the interaction between a light beam and a photonic lattice optically-induced with partially coherent light. We demonstrate a clear transition from two-dimensional discrete diffraction to discrete solitons in such a partially coherent lattice, and show that the nonlinear interaction process is associated with a host of new phenomena including lattice dislocation, lattice deformation, and creation of structures akin to optical polarons.

Closely spaced nonlinear waveguide arrays have recently been the focus of considerable attention. This is due, in part, to their strong link with the emerging science and technology of nonlinear photonic crystals and their ability to discretize light propagation [1, 2]. In such array structures, the collective behavior of wave propagation exhibits intriguing phenomena that are also encountered in many other discrete nonlinear systems, for example in biology, solid state physics and Bose-Einstein condensates [3-5]. In optics, the diffraction dynamics of a light beam is profoundly affected even in linear waveguide lattices due to evanescent coupling between nearby waveguide sites, leading to discrete diffraction. If the waveguide array is embedded in a nonlinear medium, more interesting wave behavior is expected to occur. For instance, a balance between discrete diffraction and nonlinear self-focusing can lead to optical self-trapped states better known as discrete solitons (DS) [6, 7]. In nonlinear optics, DS were predicted in 1988 [1], and were demonstrated in one-dimensional (1D) AlGaAs semiconductor waveguide arrays ten years later [6]. Recently, a theoretical study suggested that DS could also form in optically-induced waveguide arrays [8]. This soon led to the experimental observation of two-dimensional (2D) DS in such waveguide arrays induced via coherent beam interference [9]. Meanwhile, pixel-like spatial solitons have also been successfully employed to establish stable 2D nonlinear photonic lattices in photorefractive crystals [10, 11]. In this latter configuration, the lattice itself experiences a strong nonlinearity, and as a result, it becomes considerably more susceptible to modulation instability (MI) and soliton-induced deformation. This in turn brings about the interesting possibility of studying optical soliton-lattice interactions that might exhibit many of the basic characteristic features of other physical processes such as those encountered in polaron excitation/formation in solid state physics [12].

In this Letter, we report on experimental demonstration of 2D discrete solitons and soliton-induced dislocation/deformation in photonic lattices created by partially *incoherent* light. By exploiting the anisotropic properties of the photorefractive nonlinearity, the waveguide lattices can be conveniently operated in either the *linear* or *nonlinear* regime [8]. In the *nonlinear* regime, by launching a soliton (probe) beam into such a lattice, we observe transverse velocity slow-down of the probe as well as soliton-induced lattice dislocation, and creation of optical structures that are analogous to polarons in solid state physics. Conversely, when the lattice is operated in the *linear* regime, so that itself does not experience any strong nonlinearity during propagation, we observe that the probe beam evolves from discrete diffraction to DS as the level of nonlinearity for the probe is increased. We emphasize that, different from previous experiments in which the lattice was created by *coherent* multi-beam interference [9], the DS reported here are hosted in a partially *incoherent* photonic lattice. This in turn enables, through coherence control, stable lattice formation due to suppression of incoherent MI [13]. In fact, it is in such a stable lattice that detailed features of transition from 2D discrete diffraction to formation of DS are clearly demonstrated. Our experimental results are in good agreement with the theoretical analysis of these effects.

The experimental setup for our study is similar to those used in Refs [10, 14]. A partially spatially incoherent beam ($\lambda$=488 nm) is created with a rotating diffuser, and a biased photorefractive crystal (SBN:60, 5 x 5 x 8 mm$^3$, $r_{33}$=280 pm/V and $r_{13}$= 24 pm/V) is employed to provide a self-focusing noninstantaneous nonlinearity, as in previous demonstration of incoherent solitons [15]. To generate a 2D-waveguide lattice, we use an amplitude mask to spatially modulate the otherwise uniform incoherent beam after the diffuser. The mask is then imaged onto the input face of the crystal, thus creating a pixel-like input intensity pattern [10]. A Gaussian beam split from the same laser is used as the probe beam propagating along with the lattice. In addition, a uniform incoherent background beam is used as "dark illumination" for fine-tuning the nonlinearity [15].

In our experiment, the probe beam is extraordinarily polarized and "fully" coherent (i.e., it does not pass through the diffuser), while the lattice beam is partially coherent and its polarization can be either extraordinary (*e*) or ordinary (*o*) as needed. In general, in an anisotropic photorefractive crystal, the nonlinear index change experienced by an optical beam depends on its polarization as well as on its intensity. Under appreciable bias conditions, i.e., when the photorefractive screening nonlinearity is dominant, this index change is approximately given by $\Delta n_e = [n_e^3 r_{33} E_0 / 2](1+I)^{-1}$ and $\Delta n_o = [n_o^3 r_{13} E_0 / 2](1+I)^{-1}$ for e-polarized and o-polarized beams, respectively [8], different from standard saturable Kerr nonlinearity. Here $E_0$ is the applied electric field along the crystalline c-axis (x-direction), and *I* is the intensity of the beam normalized to the background illumination. Due to the difference between the nonlinear electro-optic coefficient $r_{33}$ and $r_{13}$, $\Delta n_e$ is more than 10 times larger than $\Delta n_o$ (in the SBN crystal we used) under the same experimental conditions. Thus, when it is e-polarized, the lattice beam experiences a *nonlinear* index change comparable to the probe beam, whereas it evolves almost *linearly* when it is o-polarized.

First, we present our experimental results on 2D discrete solitons. In this case, the lattice spacing has to be small enough to ensure that appreciable coupling between nearby lattice sites exists. To create a *stable* lattice with a small spacing, the lattice beam is chosen to be o-polarized and partially incoherent, thus it "sees" only a weak nonlinearity as compared to that experienced by the e-polarized probe beam. While the lattice remains nearly invariant as the bias field increases, it serves as a *linear* waveguide array for the probe beam. Typical experimental results are presented in Fig. 1, in which a 2D square lattice (spacing 20 μm and FWHM of each lattice site 10 μm) was first generated. A probe beam (whose intensity was 4 times weaker than that of the lattice) was then launched into one of the waveguide channels, propagating collinearly with the lattice. Due to weak coupling between closely spaced waveguides, the probe beam underwent discrete diffraction when the nonlinearity was low, whereas it formed a 2D discrete soliton at an appropriate level of high nonlinearity. The lattice itself was not considerably affected by the weak probe or the increased bias field. The first two photographs show the Gaussian-like probe beam at the crystal input [Fig. 1(a)] and its linear diffraction at the crystal output after 8 mm of propagation [Fig. 1(b)]. Discrete diffraction in the square lattice was observed at a bias field of 900 V/cm [Fig. 1(c)], clearly showing that most of the energy flows from the center towards the diagonal directions of the lattice. Even more importantly, a DS was observed at a bias field of 3000 V/cm [Fig. 1(d)], with most of energy concentrated in the center and the four neighboring sites along the principal axes of the lattice. (Animations of the process can be viewed online [16]). These experimental results are truly in agreement with expected behavior from the theory of discrete systems [8, 17].

We emphasize that, to form such a DS, a delicate balance has to be reached between waveguide coupling offered by the lattice and the self-focusing nonlinearity experienced by the probe beam through fine-tuning the experimental parameters (the lattice spacing, the intensity ratio, the bias field, etc.). A series of experiments were performed showing that a deviation of about 10% in the beam intensities or the applied field (both controlling the strength of the nonlinearity), or a deviation in input position/direction of the probe beam would hinder the DS formation. In addition, different from previous experiments [9], the DS reported here are hosted in a partially incoherent lattice, where the partial spatial coherence (coherence length ~100 μm) provides enhanced lattice stability due to overall suppression of incoherent MI of the lattice beam [13,14]. Should the lattice beam become e-polarized and/or "fully" coherent, the lattice (at this small spacing) becomes strongly distorted at higher levels of nonlinearity.

The above experimental observations are corroborated by numerical simulations. The evolution of the partially coherent lattice is described by the so-called coherent density approach [18], whereas that of the coherent probe beam by a paraxial nonlinear wave equation:

$$i\left(\frac{\partial f}{\partial z}+\boldsymbol{q}_x\frac{\partial f}{\partial x}+\boldsymbol{q}_y\frac{\partial f}{\partial y}\right)+\frac{1}{2k}\left(\frac{\partial^2 f}{\partial x^2}+\frac{\partial^2 f}{\partial y^2}\right)-\boldsymbol{b}\frac{1}{1+I_N}f=0\ , \quad (1)$$

$$i\frac{\partial u}{\partial z}+\frac{1}{2k}\left(\frac{\partial^2 u}{\partial x^2}+\frac{\partial^2 u}{\partial y^2}\right)-\boldsymbol{b}\frac{1}{1+I_N}u=0\ , \quad (2)$$

$$I_N(x,y,z)=\int_{-\infty}^{\infty}\int_{-\infty}^{\infty}\left|f(x,y,z,\boldsymbol{q}_x,\boldsymbol{q}_y)\right|^2 d\boldsymbol{q}_x d\boldsymbol{q}_y+I_C(x,y,z)\ , \quad (3)$$

where the coherent density function $f$ at the input is expressed as $f(x,y,\boldsymbol{q}_x,\boldsymbol{q}_y,z=0)=r^{1/2}G_N^{1/2}(\boldsymbol{q}_x,\boldsymbol{q}_y)\Phi_0(x,y)$, and $I_C=|u|^2$ is the intensity of the probe beam co-propagating with the lattice. In Eqs. (1-3), $I_N=I/I_b$ is the normalized total intensity with respect to the background illumination, and the integral term in Eq. (3) represents the intensity of the partially coherent lattice. The intensity ratio $r$ is defined as $r=I_{max}/I_b$, with $I_{max}$ being the initial maximum lattice intensity. $\boldsymbol{q}_x$ and $\boldsymbol{q}_y$ are the angles at which the coherent density propagates with respect to the z-axis. The nonlinear constant, $\boldsymbol{b}=k_0 E_0 n^3 r_{eff}/2$, is determined by the crystal parameters and the bias field $E_0$, where $k_0=2\boldsymbol{p}/\boldsymbol{l}$ is the wave number, $n$ is the index of the crystal, and the effective electrooptic coefficient for SBN:60 is $r_{eff}=(r_{33},r_{13})$ depending on the polarization. The function $G_N$ represents the Gaussian distribution of the angular power spectrum: $G_N(\boldsymbol{q}_x,\boldsymbol{q}_y)=(\boldsymbol{p}\boldsymbol{q}_0^2)^{-1}\exp[-(\boldsymbol{q}_x^2+\boldsymbol{q}_y^2)/\boldsymbol{q}_0^2]$, where $\boldsymbol{q}_0$ is related to the spatial coherence length through $l_c=\boldsymbol{l}/\sqrt{2\boldsymbol{p}}n\boldsymbol{q}_0$. $\Phi_0(x,y)$ is the spatial modulation function for the incoherent beam as imposed by the amplitude mask. Eqs. (1-3) were solved using a fast Fourier transform multi-beam propagation method. Figure 2 shows typical numerical results. The parameters chosen in simulation are close to those from experiments: lattice spacing 18 μm, FWHM of each lattice site 10.3 μm, and the intensity ratio between the lattice and the probe beam is 4.2. The lattice beam has a spatial coherence length of 100 μm. At a low bias field of 720 V/cm, discrete diffraction was observed (left), whereas at a high bias field of 2160 V/cm, formation of 2D DS was realized (right), in agreement with above experimental observations.

Next, we present results on the interaction of a soliton (probe) beam with a *nonlinear* lattice induced by 2D pixel-like spatial solitons. In this case, both the probe and the lattice beam are e-polarized and have the same wavelength. For such a configuration, stable solitonic lattices including manipulation of individual soliton channels at relatively large lattice spacing (70 μm or more) have been established previously either by use of a partially coherent lattice [10] or by phase engineering a coherent lattice [11]. Here we focus on novel aspects of behavior concerning soliton-lattice interactions when the input solitons are tilted or at small lattice spacing. When the probe beam is launched into one of the induced waveguides with a much weaker intensity compared to that of the lattice, it is simply guided by the waveguide channel without affecting the lattice. However, once the intensity of the probe becomes comparable to that of the lattice, the probe beam forms a soliton itself and thus plays an active role in the interaction process. Typically, when the probe beam is aimed at one of the waveguides but with a small angle relative to the propagation direction of the lattice, we observe lattice dislocation due to soliton dragging. Figure 3 shows such an example. A partially coherent lattice (lattice spacing ~ 65 μm; $l_c$ ~ 40 μm) was created at a bias field of 2400 V/cm. When the probe beam (intensity equal to that of the lattice) was launched at a shallow angle (~ $0.5^0$) either to the right [Fig. 3(a)] or to the left [Fig. 3(b)], the lattice overall remained uniform except for a dislocation created by the probe beam. Meanwhile, the probe itself retained its soliton identity, but its lateral shift at output was reduced significantly because of interaction with the lattice, indicating a slow-down in its transverse velocity. As shown in Fig. 3(d) and 3(e), the probe beam traveled 68 μm in the x-direction without the lattice (the spot far away from the center), whilst under the same conditions, it traveled only about 26 μm when interacting with the lattice (the spot close to the center). These two photographs were taken separately and then superimposed in the same figure. Such experimentally observed behavior is also evident in our numerical simulations. Figure 3(c) depicts the output pattern of the lattice when the probe beam was launched at an angle of $0.5^0$ towards +x-direction, and Fig. 3(f) shows the output intensity profile of the probe beam in the presence (solid line) or absence (dashed line) of the lattice under the same conditions. Similar numerical results were obtained when the probe is launched at the same angle towards –x-direction.

At a smaller lattice spacing, nonlinear soliton-lattice interaction leads to other interesting phenomena including strong lattice deformation and generation of polaron-like structures. When the probe beam was launched straight into one of the waveguide channels, a polaron-like induced structure was observed at various lattice spacing (40 μm, 50 μm and 60 μm), in which the probe soliton dragged towards it some of the neighboring sites while pushing away the other. Typical experimental results are illustrated in Fig. 4, for which the lattice spacing is 40 μm. From Fig. 4(a) to 4(c), the intensity of the probe soliton was increased while that of the lattice was kept unchanged. When the probe intensity was much higher than that of the lattice, the lattice structure became strongly deformed such that the site dislocations extended beyond the immediate neighborhood [Fig. 4(c)]. For comparison, Fig. 4(d) shows the corresponding restored lattice after the probe beam was removed and the crystal has reached a new steady-state under the same conditions. This observed interaction process is quite similar to that caused by a polaron in solid state physics during which an electron drags and dislocates heavy ions as it propagates through an ionic crystal [12]. Since the closest four neighboring solitons are equally spaced initially around the central one where the probe beam was launched, the observed behavior cannot simply be attributed to the anomalous interaction between photorefractive solitons in which attraction or repulsion merely depends on soliton mutual separation [19]. Instead, the observed polaron-like structure suggests that the probe beam might have induced certain degree of coherence to the neighboring lattice sites with different phase correlation through interaction.

In summary, we have successfully observed 2D discrete solitons in optically-induced partially-coherent photonic lattices along with a host of new phenomena arising from soliton-lattice interaction. Our results may pave the way towards the observation of similar phenomena in other relevant discrete nonlinear systems.

This work was supported by AFOSR, Research Corp., ARO MURI, and the Pittsburgh Supercomputing Center. We thank M. Segev, Y. Kivshar and J. Xu for discussion.

References:

[1] D. N. Christodoulides and R. I. Joseph, Opt. Lett. **13**, 794 (1988).

[2] S. Mingaleev and Y. Kivshar, Opt. & Photonics News, Vol. **13**, 48 (2002).

[3] A. S. Davydov, Biology and Quantum Mechanics (Pergamon, Oxford, 1982).

[4] A. J. Sievers and S. Takeno, Phys. Rev. Lett. **61**, 970 (1988).

[5] A. Trombettoni and A. Smerzi, Phys. Rev. Lett. **86**, 2353 (2001).

[6] H. S. Eisenberg et al., Phys. Rev. Lett. **81**, 3383 (1998);
    R. Morandotti et al., Phys. Rev. Lett. **83**, 2726 (1999).

[7] F. Lederer and Y. Siberberg, Opt. & Photonics News, Vol. **13**, 48 (2002).

[8] N. K. Efremidis et al., Phys. Rev. E **66**, 046602 (2002).

[9] J. W. Fleischer, et al., Nature **422**, 150 (2003).

[10] Z. Chen and K. MaCarthy, Opt. Lett. **27**, 2019 (2002).

[11] J. Petter et al., Opt.Lett. **28**, 438 (2003); M. Petrovic et al., arXiv:nlin.PS/0306041(2003).

[12] C. Kittel, Introduction to Solid State Physics (John Wiley & Sons, New York, 1996).

[13] M. Soljacic et al., Phys. Rev. Lett. **84**, 467 (2000).

[14] Z. Chen et al., Phys. Rev. E **66**, 066601 (2002).

[15] M. Mitchell et al.,Phys. Rev. Lett. **77**, 490 (1996); Z. Chen et al., Science **280**, 889 (1998).

[16] Animations of DS formation as obtained from experiment can be viewed at
    www.physics.sfsu.edu/~laser/.

[17] J. Hudock et al., "Elliptical discrete solitons in waveguide arrays", Opt. Lett., submitted.

[18] D. N. Christodoulides et al., Phys. Rev. Lett.**78**, 646 (1997).

[19] W. Krolikowski et al., Phys. Rev. Lett. **80** 3240 (1998).

Figure Captions:

Fig. 1: Experimental demonstration of a discrete soliton in a partially coherent lattice. (a) Input, (b) diffraction output without the lattice, (c) discrete diffraction at 900 V/cm, and (d) discrete soliton at 3000 V/cm. Top: 3D intensity plots; Bottom: 2D transverse patterns.

Fig. 2: Numerical results corresponding to Fig. 1(c) and Fig. 1(d). Inserts are 2D transverse patterns.

Fig. 3: Soliton-induced lattice dislocation when the probe beam was launched towards the right (a) and the left (b) at $0.5^o$. (d) and (e) show a slow-down in the transverse velocity of the probe beam. (c) and (f) are numerical results corresponding to (a) and (d), respectively.

Fig. 4: Soliton-induced polaron-like structures (a-c) and the restored lattice after the probe beam is turned off (d). From (a) to (c), the intensity ratio between the probe soliton and the lattice is 0.6, 1.0, and 1.5.

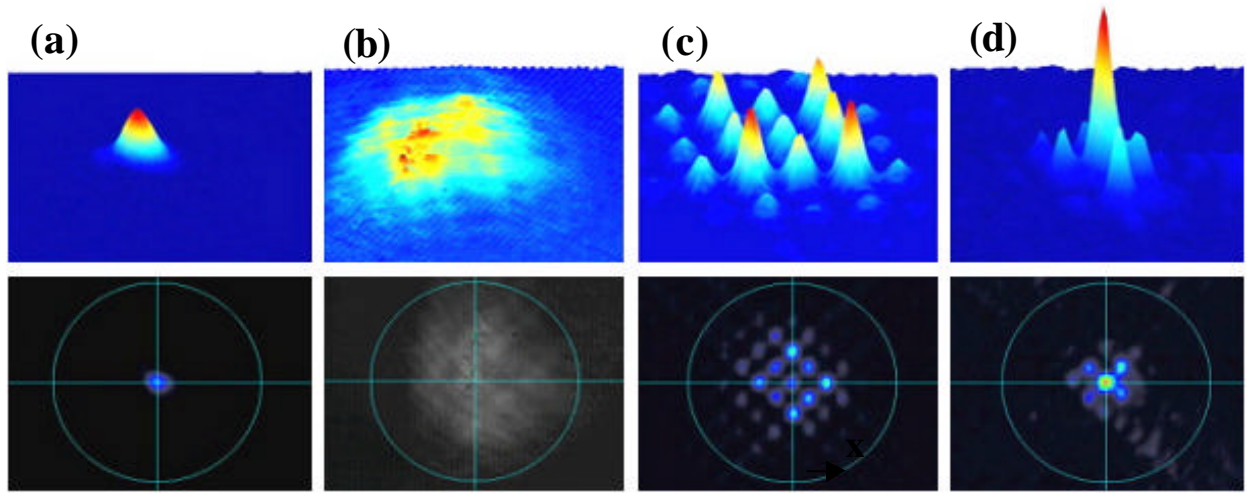

**Martin et al. Fig. 1**

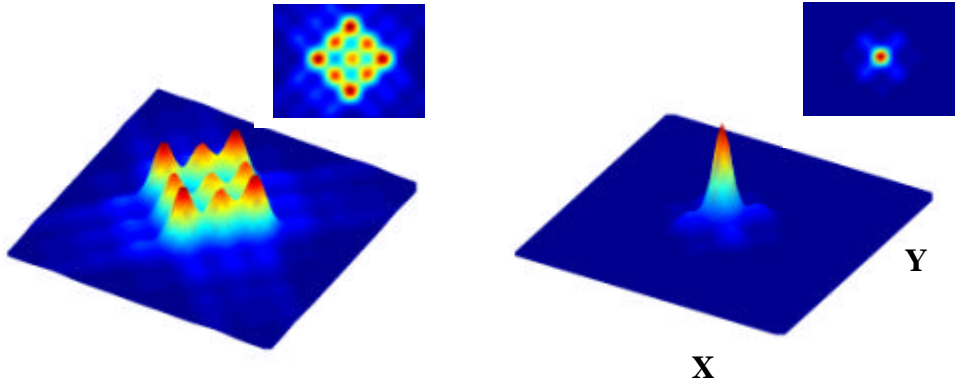

**Martin et al., Fig. 2**

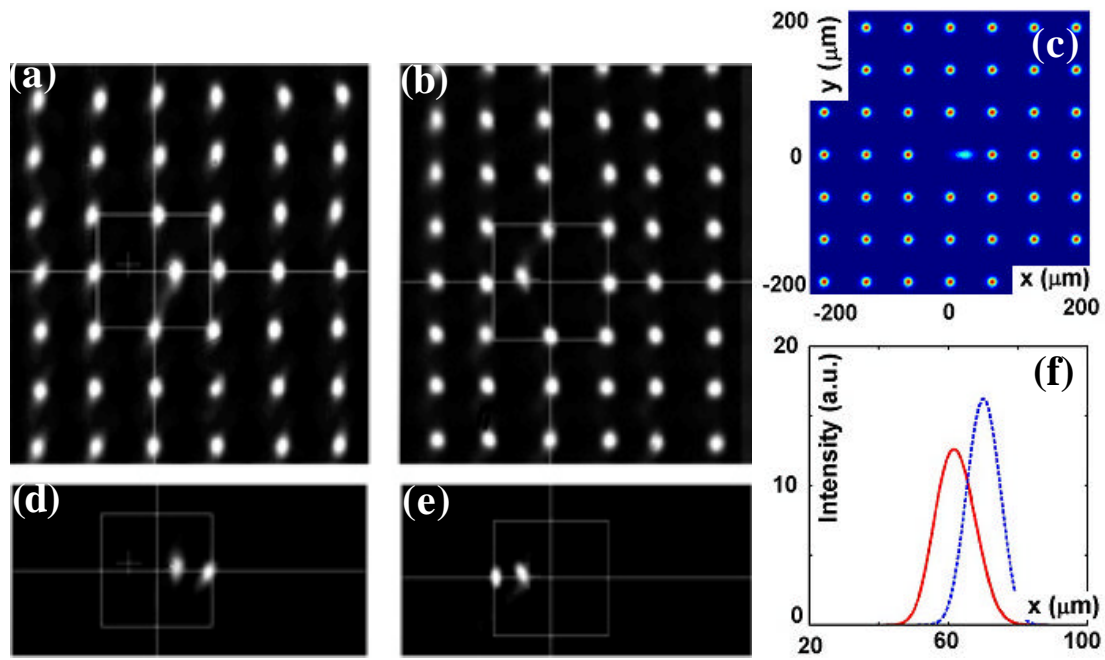

Martin et al., Fig. 3

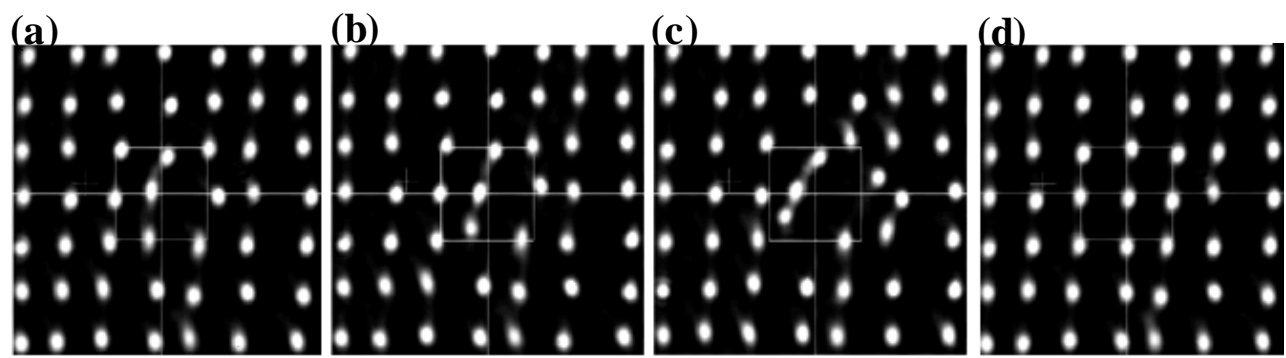

Martin et al., Fig. 4